\documentclass[10pt]{article}

\usepackage{amsmath, amssymb, latexsym, cite}
\usepackage[dvips]{graphicx}

\newcommand{\TeV}{{\text{TeV}}}
\newcommand{\GeV}{ \text{GeV} }

\begin{document}

%%%%%%%%%%%%%%%%%%%%%%%%%%%%%%%%%%%%%%%%%%%%%%%%%%%%%%
\begin{titlepage}
\begin{center}

{\small
\hfill IPMU12-0018 }

\vspace{2.0cm}

{\large\bf Comprehensive Analysis on the Light Higgs Boson Scenario}

\vspace{2.0cm}

{\bf Masaki Asano}$^{\it (a, b, c)}$, 
{\bf Shigeki Matsumoto}$^{\it (d)}$,
{\bf Masato Senami}$^{\it (e),}
$\footnote{This author is now in Department of Micro Engineering, Kyoto University, Kyoto 606-8501, Japan}, \\
and
{\bf Hiroaki Sugiyama}$^{\it (f)}$

\vspace{1.5cm}

{\it $^{(a)}${II. Institute for Theoretical Physics, University of Hamburg,\\
Luruper Chausse 149, DE-22761 Hamburg, Germany}}\\
\vspace{0.1cm}
{\it $^{(b)}${Department of Physics, University of Tokyo, Tokyo 113-0033, Japan}}\\
\vspace{0.1cm}
{\it $^{(c)}${Department of Physics, Tohoku University, Sendai 980-8578, Japan}}\\
\vspace{0.1cm}
{\it $^{(d)}${IPMU, TODIAS, University of Tokyo, Kashiwa 277-8583, Japan}}\\
\vspace{0.1cm}
{\it $^{(e)}${ICRR, University of Tokyo, Kashiwa 277-8582, Japan}}\\
\vspace{0.1cm}
{\it $^{(f)}${Department of Physics, Ritsumeikan University, Kusatsu 525-8577, Japan}}\\

\vspace{2.0cm}

\abstract{Comprehensive analysis on the light Higgs scenario of the minimal supersymmetric standard model is performed in the framework of the non-universal Higgs mass model~(NUHM). The NUHM is known to be the simplest framework realizing the light Higgs scenario with the unifications of gaugino and sfermion masses at the scale of grand unified theory being consistent. All important constraints from collider experiments, flavor-changing processes, and cosmological observations are considered in order to clarify the allowed region of the model parameter space, where the Markov Chain Monte Carlo method is used to obtain the region. We show that the region is excluded by current LHC results for the SUSY particle searches. We discuss possible extensions to save the LHS\@. Even with such extensions, the measurement of $B_s \to \mu^+\mu^-$ would enable us to test the LHS in the near future\@.}

\end{center}
\end{titlepage}

%%%%%%%%%%%%%%%%%%%%%%%%%%%%%%%%%%%%%%%%%%%%%%%%%%%%%%
\setcounter{footnote}{0}

\section{Introduction}
\label{sec:intro}

In spite of tremendous successes of the standard model~(SM), both hierarchy and dark matter~(DM) problems require us to consider new physics beyond the SM\@. The hierarchy problem is essentially given rise to by quadratically divergent corrections to the Higgs mass term, so that the new physics is expected to appear at around or less than ${\cal O}(1)\,\TeV$ scale. On the other hand, recent cosmological observations such as the Wilkinson Microwave Anisotropy Probe~(WMAP) experiment~\cite{WMAP,recentWMAP} have established the existence of dark matter, which cannot be accounted for in the framework of the SM\@. Among several candidates for dark matter, the weakly interacting massive particle~(WIMP) is one of the most promising candidates~\cite{DMreview}. Since the mass of the WIMP is predicted to be less than ${\cal O}(1)\,\TeV$ scale in order for its correct abundance observed today~\cite{WMAP,recentWMAP}, the WIMP is also likely to be obtained by the new physics at the TeV scale.

The minimal supersymmetric standard model~(MSSM)~\cite{BookDrees} is thought to be the most promising candidate for the new physics. It gives solutions to both hierarchy and dark matter problems simultaneously and enables us to unify SM gauge interactions. The direct search of the Higgs boson at the LEP experiment~\cite{LEPSM} severely constrains the parameter space of the MSSM, because the model predicts the mass of the lightest neutral Higgs boson $h$ to be the same as or less than the $Z$ boson mass at the tree level. In principle, there are two prescriptions to avoid the constraint. One is that $h$ is composed mainly of the up-type Higgs. Radiative corrections from scalar top quarks then push up the mass of $h$~\cite{radiative}. This scenario has frequently been discussed so far. In the limit where superpartners are very heavy (although it needs fine-tunings), signals of the Higgs boson at collider experiments becomes very similar to that of the SM\@.

Another prescription is that $h$ is composed mainly of the down-type Higgs. In this case, there are regions to evade the constraint though radiative corrections from scalar bottom quarks do not push up the $h$ mass enough. The LEP constraint is evaded because the coupling of the $ZZh$ interaction can be suppressed~\cite{LHSKane, LHSBelyaev, LHSDrees, LHSKim, Asano:2007gv, LEPSUSY, Asano:2009kj, LHSdummy}. This is called the light Higgs boson scenario~(LHS). The LHS can be tested at the LHC experiment because the difference from the SM inevitably exists; for example, the masses of other Higgs bosons in the LHS as well as that of $h$ are predicted to be around the $Z$ boson mass. The LHS has some attractive features. The LEP anomaly~\cite{LHSDrees}, which is the excess of the Higgs-like events with the Higgs mass being around $98\,\GeV$ reported by the LEP collaborations, can be explained. Severe fine-tunings on the supersymmetric~(SUSY) little hierarchy problem is not required~\cite{LHSKim}. It is also worth noting that the LHS is consistent with a simple framework of grand unified theory~(GUT) and seesaw mechanism~\cite{Asano:2008gi}.
The cosmological aspects have also studied~\cite{ Asano:2007gv,Asano:2009kj,Kim:2008uh,Funakubo:2009eg}.
Several recent papers have studied the expected parameter
space in MSSM~\cite{Eriksson:2008cx}.

In the LHS, the charged Higgs boson significantly contributes to the $b \to s \gamma$ process, since their mass is very light as that of $Z$ boson. This contribution turns out to be too large, because $m_{H^+} > 350\,\GeV$ is required for the type~II two Higgs doublet model~\cite{Gambino:2001ew}. This should be hence canceled out with SUSY contributions. Because of this reason, the LHS predicts that both chargino and stop are light enough. This fact inevitably leads to large Br$(B_s \to \mu^+ \mu^-)$. As a result, the LHS must have large SUSY contribution to $b \to s \gamma$, while small contribution to $B_s \to \mu^+ \mu^-$. In order to calculate these processes accurately, the LHS should be studied comprehensively in the framework of a unification model. In this paper, we analyze the LHS in the framework of the non-universal Higgs masses model~(NUHM), which is the simplest model realizing the LHS in the MSSM\@. Using the Markov Chain Monte Carlo~(MCMC) method, we search the parameter region consistent with all important constraints from collider experiments, flavor-changing processes, and cosmological observations. In particular, we impose the constraint of the dark matter relic abundance observed by WMAP~\cite{WMAP} and the bound by direct detection experiments~\cite{Aprile:2011hi}. We find that the LHS in the framework of the NUHM is marginally inconsistent with SUSY particle search in the LHC experiments~\cite{SUSY-ATLAS,Chatrchyan:2011zy}.

This paper is organized as follows. In the next section, we briefly introduce the LHS in the framework of the MSSM, and also explain how the NUHM realizes the LHS\@. In Section~\ref{sec:sim}, the framework of our analysis for the comprehensive study of the LHS is shown in addition to some detailed explanations of experimental constrains used in our analysis. All of our physics results is presented in Section~\ref{sec:results}. Section \ref{sec:summary} is devoted to summary. We also discuss some relaxations of experimental constraints and the boundary condition of the NUHM\@.

\section{Light Higgs boson scenario}
\label{sec:LHS}

The MSSM predicts two Higgs doublets denoted by $H_u$ and $H_d$. One Higgs doublet $H_u$ gives the masses of up-type quarks, while down-type quarks as well as charged leptons acquire their masses by interactions with $H_d$. Five (real) scalar components in $H_u$ and $H_d$ remain as physical states; one CP-odd pseudo-scalar Higgs boson $A$, a pair of charged Higgs bosons $H^\pm$, and two CP-even Higgs bosons $h$ (lighter) and $H$ (heavier). In the basis of (${\rm Re} \, H_d^0, {\rm Re} \, H_u^0$) with $H_d^0$ and $H_u^0$ being neutral components of the two Higgs doublets, the mass matrix is given by
\begin{eqnarray}
\left(
\begin{array}{cc}
m_A^2 s_\beta^2 + m_Z^2 c_\beta^2 + \Delta_{dd} &
- ( m_A^2 + m_Z^2 ) s_\beta c_\beta + \Delta_{du} \\ 
- ( m_A^2 + m_Z^2 ) s_\beta c_\beta + \Delta_{du} &
m_A^2 c_\beta^2 + m_Z^2 s_\beta^2 + \Delta_{uu}
\end{array}
\right).
\end{eqnarray}
Here, $m_A$ ($m_Z$) is the mass of the pseudo-scalar Higgs boson (the $Z$ boson) and $c_\beta$ ($s_\beta$) $\equiv$ $\cos\beta$ ($\sin\beta$), where the ratio of two vacuum expectation values $\langle H_u \rangle$ and $\langle H_d \rangle$ defines the angle $\beta$ as $\tan \beta \equiv \langle H_u \rangle/\langle H_d \rangle$. Radiative corrections in each component of the mass matrix are summarized in $\Delta_{ij}$, and their detailed expressions are found in Ref.~\cite{BookDrees}. The mass eigenstate of CP-even Higgs bosons $h$ and $H$ are obtained by diagonalizing this matrix. With the use of the mixing angle $\alpha$, the mixing matrix to diagonalize the mass matrix is given by
\begin{eqnarray}
\left( \begin{array}{c} h \\ H \end{array} \right) =
\left(
\begin{array}{cc}
- \sin\alpha & \cos\alpha \\
\cos\alpha &  \sin\alpha
\end{array}
\right)
\left(
\begin{array}{c} {\rm Re} \, H_d^0  \\ {\rm Re} \, H_u^0 \end{array}
\right).
\end{eqnarray}

The LEP collaborations have given constraints on masses and couplings of the Higgs bosons not only in the SM but also in the MSSM~\cite{LEPSUSY}. According to their results, the lightest Higgs boson $h$ in the MSSM can be lighter than 114.4\,\GeV, which is the lower bound on the SM Higgs boson, when the coupling between $h$ and $Z$ bosons $g_{ZZh}$ is suppressed enough. This is because the discovery of the Higgs boson at the LEP2 experiment relied on the process $e^+ e^- \to Zh$ through the s-channel exchange of the $Z$ boson. In the MSSM, this coupling is given by $g_{ZZh} = g_{ZZh}^{\rm (SM)} \sin(\beta - \alpha)$ with $g_{ZZh}^{\rm (SM)}$ being the coupling between $Z$ bosons and the Higgs boson of the SM\@. In order to obtain the Higgs boson $h$ lighter than 114.4\,\GeV, a suppressed $\sin(\beta - \alpha)$ is required. Since $\tan\beta$ is expected to be large enough to satisfy experimental constraints as shown in the following sections, a large mixing angle $\alpha \sim \pi/2$ is needed. On the other hand, the LEP bound on the SM Higgs boson is also applied to the heavier Higgs boson $H$. This constraint is complementary to that on $h$, because the coupling between $H$ and $Z$ bosons is given by $g_{ZZH} = g_{ZZh}^{\rm (SM)} \cos(\beta - \alpha)$. Furthermore, the coupling between $Z$, $A$, and $h$ is also proportional to $\cos(\beta - \alpha)$, so that the search of Higgs bosons at the LEP2 experiment using the $e^+ e^- \to Ah$ process should be paid attention to. Since the $g_{ZAh}$ coupling originates in a derivative coupling, this process does not provide a serious constraint due to P-wave suppression when $m_h > 90\,\GeV$~\cite{LEPSUSY}.

Since the constrained MSSM~(CMSSM)~\cite{CMSSM,Kane:1993td} is too restricted to represent all degrees of freedom of the Higgs sector, the LHS is impossible to be realized. Particularly, the soft SUSY breaking masses of two Higgs doublet fields $m_{H_u}^2$ and $m_{H_d}^2$ are fixed by other parameters such as universal supersymmetric particle masses $m_0$ and $m_{1/2}$. As a result, when $m_h$ is lighter than 114.4\,\GeV, all other supersymmetric particles become also very light, and such a light Higgs boson has already been excluded. The mass spectrum of the Higgs bosons in the CMSSM is therefore inevitably the decoupled type, namely, the lightest Higgs boson is up-type (${\rm Re} \, H_u^0$, meaning $\alpha \sim 0$) with the mass of the order of $m_Z$, while other Higgs bosons are much heavier, of the order of 1\,\TeV. This fact means that constraints on $m_{H_u}^2$ and $m_{H_d}^2$ should be relaxed, or even be chosen freely from other sparticle masses in order to realize the LHS\@. One interesting possibility is to relax the boundary condition $m_{H_u}^2 = m_{H_d}^2 = m_0^2$ to be $m_{H_u(H_d)}^2 = (1 + \delta_{H_u(H_d)}) m_0^2$ at the GUT scale $M_G$. Since Higgs multiplets are not necessarily the same ones of other matter superfields, these relaxed boundary conditions do not break the simple framework of GUTs. The minimal extension to the CMSSM with this relaxation is called the NUHM\@. This model is hence suitable for the search of the parameter space of the LHS as a reference model. The NUHM has six model parameters, $(m_0, \, m_{1/2}, \, A_0 , \, \tan\beta, \, \mu, \, m_A )$, where $(m_0, \, m_{1/2}, \, A_0)$ are defined at $M_G$ and others are defined at the electroweak scale. Parameter degrees of freedom for $\mu$ and $m_A$ are translated into $m_{H_u}$ and $m_{H_d}$ at $M_G$. This parameterization allows us to deal with the masses of two Higgs doublets as free parameters.

\section{Framework of our analysis}
\label{sec:sim}

\subsection{Markov Chain Monte Carlo}

We use the MCMC method to clarify the allowed region of the LHS\@. The MCMC method is a random sampling algorithm, which gives a series of parameter sets called the Markov chain as output~\cite{recipes}. The samples of the chain obey the distribution which is proportional to a given distribution function. The distribution function of our interest is a posterior probability distribution function of model parameters $x$ under experimental data $D$. Bayes' theorem tells us that the posterior probability distribution function $P(x|D)$ satisfies the following equation,
\begin{eqnarray}
  P(x|D) = \frac{ P(D|x) P(x) }{ \sum_{x^\prime} P(D|x^\prime) P(x^\prime) },
\end{eqnarray}
where $P(x)$ is the prior probability function reflecting our knowledge about model parameters $x$, while $P(D|x)$ represents the likelihood for the distribution function of experimental data $D$ at given model parameters $x$. In our analysis, a linearly flat prior has been used for $P(x)$, where $P(x)dx$ gives a constant probability. For experimental data $D$, we have used constraints shown in the upper part of Table~\ref{tab:constraints}. Constraints on $\Omega_{\rm DM} h^2$ and ${\rm Br}(b\to s\gamma)$ are assumed to have Gaussian distributions. Other constraints are adopted as boundaries of model parameters. The last two constraints in Table~\ref{tab:constraints} are applied later to samples generated by the MCMC\@.

\begin{table}[t]
\begin{center}
\begin{tabular}{|l|l|c|}
\hline
{}
 & Constraints
 & References \\
\hline
%%%%%%%%%%%%%%%%%%%%%%%%%
Upper bound on $g_{ZZh}$ ($g_{ZZH}$)
 & See the text
 & \cite{LEPSUSY} \\
%%%%%%%%%%%%%%%%%%%%%%%%%
Upper bound on $g_{ZAh}$ ($g_{ZAH}$)
 & See the text
 & \cite{LEPSUSY} \\
%%%%%%%%%%%%%%%%%%%%%%%%%
Relic abundance of dark matter ($\Omega_{\rm DM}h^2$)
 & $0.1099 \pm 0.0062$
 & \cite{WMAP} \\
%%%%%%%%%%%%%%%%%%%%%%%%%
Direct detection of dark matter
 & See the text
 & \cite{Aprile:2011hi} \\
%%%%%%%%%%%%%%%%%%%%%%%%%
Anomalous magnetic moment of muon ($\Delta a_\mu$)
 & $9.5 < (\Delta a_\mu \times 10^{10}) < 41.5$
 & \cite{Nakamura:2010zzi} \\
%%%%%%%%%%%%%%%%%%%%%%%%%
${\rm Br}(b\to s\gamma)$
 & $(3.52 \pm 0.25) \times 10^{-4}$
 & \cite{Barberio:2008fa} \\
%%%%%%%%%%%%%%%%%%%%%%%%%
${\rm Br}(B_s\to \mu^+ \mu^-)$
 & $< 1.08\times 10^{-8}$
 & \cite{Bsmumu-LHC} \\
%%%%%%%%%%%%%%%%%%%%%%%%%%
${\rm Br}(B \to \tau \nu)$
 & $0.68 < r^{\text{MSSM}} < 2.76$
 & \cite{Barberio:2008fa, Akeroyd:2010qy} \\
%%%%%%%%%%%%%%%%%%%%%%%%%
The lightest neutralino mass
 & $> 50.3 \, \GeV$
 & \cite{LEPSUSYweb} \\
%%%%%%%%%%%%%%%%%%%%%%%%%
The lightest chargino mass
 & $> 103.5 \, \GeV \ (92.4\,\GeV)$
 & \cite{LEPSUSYweb} \\
%%%%%%%%%%%%%%%%%%%%%%%%%
Right-handed selectron mass
 & $> 99.9 \, \GeV \ (73\,\GeV)$
 & \cite{LEPSUSYweb} \\
%%%%%%%%%%%%%%%%%%%%%%%%%
Right-handed smuon mass
 & $> 94.9 \, \GeV \ (73\,\GeV)$
 & \cite{LEPSUSYweb} \\
%%%%%%%%%%%%%%%%%%%%%%%%%
Right-handed stau mass
 & $> 86.6 \, \GeV \ (73\,\GeV)$
 & \cite{LEPSUSYweb} \\
%%%%%%%%%%%%%%%%%%%%%%%%%
Sneutrino masses
 & $> 94 \, \GeV \ (43\,\GeV)$
 & \cite{Abdallah:2003xe} \\
%%%%%%%%%%%%%%%%%%%%%%%%%
Stop mass
 & $> 95 \, \GeV \ (65\,\GeV)$
 & \cite{LEPSUSYweb} \\
%%%%%%%%%%%%%%%%%%%%%%%%%
Sbottom masses
 & $> 95 \, \GeV \ (59\,\GeV)$
 & \cite{LEPSUSYweb} \\
%%%%%%%%%%%%%%%%%%%%%%%%%
Squark masses (1st and 2nd generations)
 & $> 379 \, \GeV$
 & \cite{:2007ww}\\
%%%%%%%%%%%%%%%%%%%%%%%%%
Gluino masses
 & $> 308 \, \GeV$
 & \cite{:2007ww}\\
\hline
%%%%%%%%%%%%%%%%%%%%%%%%%
Squark masses (1st and 2nd generations)
 & $> 1100\,\GeV$
 & \cite{SUSY-ATLAS,Chatrchyan:2011zy} \\
%%%%%%%%%%%%%%%%%%%%%%%%%
Gluino masses
 & $> 750\,\GeV$
 & \cite{SUSY-ATLAS,Chatrchyan:2011zy} \\
\hline
\end{tabular}
\caption{\small
Experimental constraints used in our comprehensive analyses. See refs.~\cite{recentWMAP}, \cite{Hagiwara:2011af}, and \cite{Asner:2010qj} for updated constraints on $\Omega_{\text{DM}} h^2$, $\Delta a_\mu$, and ${\rm Br}(b\to s\gamma)$, respectively. The last two constraints shown in the table, which are obtained at the LHC experiment, are used later for the samples generated by the MCMC\@. Values in parentheses are applied only for limited cases. See the text for more details.}
\label{tab:constraints}
\end{center}
\end{table}

%%%%%%%%%%%%%%%%%%%%%%%%%%%%%%%
%%%  subsec: constraints  %%%%%
%%%%%%%%%%%%%%%%%%%%%%%%%%%%%%%
\subsection{Constraints used in the MCMC calculation}
\label{sec:constraints}

Here, we explain some details of constraints used in our MCMC study. In addition to experimental constraints shown in Table~\ref{tab:constraints}, we also take a theoretical constraint into account in our study, which relates to the stability of the vacuum we are living. Constraint from the $\rho$ parameter is not imposed because we found that it is always satisfied very well by our MCMC samples of the LHS where Higgs bosons have similar masses. \\

%%%  constraint from ZZh and ZZH in LEP  %%%%%
\noindent
{\bf $\bullet$ Constraints on the $ZZh$ and $ZZH$ coupling constants} \\

\noindent
The upper bound on $(g_{ZZ{\mathcal H}_1}^{}/g_{ZZh}^{(SM)})^2$ is obtained by the search for the $e^+ e^- \to Z {\mathcal H}_1$ process in the LEP2 experiment~\cite{LEPSUSY}. This corresponds to the upper bounds on $\sin^2(\beta-\alpha)$ for ${\mathcal H}_1 = h$ and $\cos^2(\beta-\alpha)$ for ${\mathcal H}_1 = H$ in the MSSM\@. The bound at 95\%~C.L.\ is shown in Table~14 of Ref.~\cite{LEPSUSY} as a function of the ${\mathcal H}_1$ mass. Bounds ${\rm UB}_b$ and ${\rm UB}_\tau$ are available for the cases of ${\rm Br}( {\mathcal H}_1 \to b \bar{b})=100\,\%$ and ${\rm Br}( {\mathcal H}_1 \to \tau \bar{\tau})=100\,\%$, respectively. For arbitrary values of ${\rm Br}( {\mathcal H}_1 \to b \bar{b})$ and ${\rm Br}( {\mathcal H}_1 \to \tau \bar{\tau})$, we use the upper bound ${\rm UB}_{\rm tot} ({\mathcal H}_1)$ defined by
\begin{eqnarray}
{\rm UB}_{\rm tot} ({\mathcal H}_1)
=
\left(
\frac{ {\rm Br}( {\mathcal H}_1 \to b \bar b)^2 }{ {\rm UB}_b^2 }
+
\frac{ {\rm Br}( {\mathcal H}_1 \to \tau \bar \tau)^2 }{ {\rm UB}_\tau^2 }
\right)^{-1/2}.
\end{eqnarray}
In order to satisfy the constraint for the lightest Higgs boson $h$, too small $\sin(\beta - \alpha)$ is not required in the range $90\,GeV < m_h < 114.4\,GeV$, and, in fact,  $\sin(\beta - \alpha) \lesssim 0.5$ is enough. On the other hand, $h$ can evade the LEP2 constraint even for $m_h < 90\,\GeV$ when $\sin (\beta - \alpha)$ is smaller than 0.2. The constraint is so tight that such a region is very restricted as mentioned in the next section. It is to be noted that another coupling $g_{ZZH} \propto \cos(\beta - \alpha)$ inevitably be large when the coupling $g_{ZZh} \propto \sin(\beta - \alpha)$ is small. Hence, the heavier Higgs boson $H$ tends to receive a severe constraint from the LEP2 experiment in the LHS\@. \\

%%%  constraint from ZAH and ZAH in LEP  %%%%%
\noindent
{\bf $\bullet$ Constraints on the $ZAh$ and $ZAH $ coupling constants} \\

\noindent
Upper bounds on the MSSM suppression factors ($\cos^2(\beta-\alpha)$ for ${\mathcal H}_1 = h$ and $\sin^2(\beta-\alpha)$ for ${\mathcal H}_1 = H$) are obtained also by the search for $e^+ e^- \to A {\mathcal H}_1$ in the LEP2 experiment, which are shown in Table~17 of Ref.~\cite{LEPSUSY}. Three bounds (we call ${\rm UB}_{bb}$, ${\rm UB}_{b\tau}$, and ${\rm UB}_{\tau\tau}$) are presented there. Bounds ${\rm UB}_{bb}$ and ${\rm UB}_{\tau\tau}$ are obtained for cases where both of Higgs bosons decay only into $b\bar{b}$ and $\tau\bar{\tau}$, respectively. The bound ${\rm UB}_{b\tau}$ is given for the case where one of Higgs bosons decays into $b\bar{b}$ and the other does into $\tau\bar{\tau}$. For arbitrary branching ratios of ${\mathcal H}_1$ and $A$ decays, we constrain $\cos^2(\beta-\alpha)$ and $\sin^2(\beta-\alpha)$ by using ${\rm UB}_{\rm tot}({\mathcal H}_1, A)$ defined by
\begin{eqnarray}
{\rm UB}_{\rm tot}({\mathcal H}_1, A)
&=&
\left( 
 \frac{ {\rm Br} ({\mathcal H}_1 \to \tau \bar \tau)^2
        {\rm Br} (A \to \tau \bar \tau)^2 }
      { {\rm UB}_{\tau\tau}^2 }
 +
 \frac{ {\rm Br} ({\mathcal H}_1 \to b \bar b)^2
        {\rm Br} (A \to b \bar b)^2 }
        { {\rm UB}_{bb}^2 } 
\right.
\nonumber \\
&&
\left.
 +
 \frac{ {\rm Br} ({\mathcal H}_1 \to \tau \bar \tau)^2
        {\rm Br} (A \to b \bar b)^2 
        +
        {\rm Br} ({\mathcal H}_1 \to b \bar b)^2
        {\rm Br} (A \to \tau \bar \tau)^2 }
        { {\rm UB}_{b\tau}^2 }
\right)^{-1/2} .
\end{eqnarray}
Since the upper bounds come originally from the production cross section of two scalars with the center of mass energy of $\sim 200\,\GeV$, we do not have a severe constraint on the coupling due to the P-wave suppression when $m_A + m_{{\mathcal H}_1} \sim 200\,\GeV$. Roughly speaking, the constraint gives a lower bound on $m_A$ as $m_A \gtrsim (200\,\GeV - m_h)$. \\ %as can be seen in the next section. \\

%%%%%%%%%  constraint from DM abundance  %%%%%%%%%
\noindent
{\bf $\bullet$ Constraint on the dark matter abundance} \\

\noindent
The lightest neutralino plays the role of the DM in the NUHM, and its relic abundance $\Omega_{\rm DM}h^2$ is computed by solving the Boltzmann equation numerically. This abundance is constrained by the WMAP experiment~\cite{WMAP, recentWMAP}. We have used the five-year result~\cite{WMAP} of the experiment ($\Omega_{\text{DM}} h^2 = 0.1099 \pm 0.0062$) as shown in Table~\ref{tab:constraints}, which is obtained by the WMAP data alone and seems conservative. Recent result is found in Ref.~\cite{recentWMAP} as $\Omega_{\text{DM}} h^2 = 0.1120 \pm 0.0056$. The relic abundance is calculated with the use of the micrOMEGAs~2.2.CPC code~\cite{Belanger:2008sj} which allows us to deal with resonant and coannihilation processes appropriately. \\

%%%%%%%%%  constraint from direct detection of DM  %%%%%%%%%
\noindent
{\bf $\bullet$ Constraints from direct detections of dark matter} \\

\noindent
Since all Higgs bosons are predicted to be light in the LHS, the spin-independent scattering cross section between the dark matter and a nucleon (a nucleus) is expected to be large. Direct detection experiments of the dark matter therefore give an important constraint on model parameters. The most severe constraint is now given by the XENON~100 experiment~\cite{Aprile:2011hi}. We have used the upper bound on the cross section as a function of the dark matter mass in our MCMC calculation. This constraint implicitly assumes that the local dark matter density around us is 0.3\,\GeV/cm$^3$. 
This constraint is relaxed if we consider a non-smooth distribution of dark matter in our halo. We take into account this uncertainty on the dark matter density by relaxing the constraint of XENON~100 by factor 2~\cite{Kamionkowski:2008vw}.
In our work, we set $y=0$ for the strangeness component in a nucleon~\cite{Ohki:2008ff}.\\

%%%%%%%%% constraint from muon g-2  %%%%%%%%%
\noindent
{\bf $\bullet$ Constraint on $g_\mu - 2$} \\

\noindent
The experimental result of the anomalous magnetic moment of muon is known to be deviated from the SM prediction. The deviation has been reported as $\Delta a_\mu = (25.5 \pm 8.0 )\times 10^{-10}$~\cite{Nakamura:2010zzi} (See also Ref.~\cite{Hagiwara:2011af}), and we use this value with the $2\sigma$ region, $9.5 < (\Delta a_\mu \times 10^{10}) < 41.5$, in order to constrain model parameters. \\

%%%%%%%%%  constraint from BR(b to s gamma)  %%%%%%%%%
\noindent
{\bf $\bullet$ Constraint on the $b \to s \gamma$ process} \\

\noindent
In our MCMC calculation, we have used ${\rm Br}(b \to s \gamma) = (3.52 \pm 0.25)\times 10^{-4}$~\cite{Barberio:2008fa}. Recent experimental bound, ${\rm Br}(b \to s \gamma) = (3.55 \pm 0.24 \pm 0.09)\times 10^{-4}$, is found in Ref.~\cite{Asner:2010qj}. The SM prediction to the branching ratio is, on the other hand, ${\rm Br}(b \to s \gamma) = (2.98 \pm 0.26) \times 10^{-4}$~\cite{Becher:2006pu} which is somewhat smaller than the observed value. Interestingly, the LHS is possible to resolve the discrepancy, because the contribution of ${H^\pm}$ makes ${\rm Br}(b \to s \gamma)$ increased from the SM prediction~\cite{Gambino:2001ew}. Since all Higgs bosons are as light as the $Z$ boson mass in the LHS, $H^\pm$ contributes to ${\rm Br}(b \to s \gamma$) significantly and this contribution is, in fact, even too large. Actually, in the type~II two Higgs doublet model, $m_{H^+} > 350\,\GeV$ is required~\cite{Gambino:2001ew}. Hence, it should be canceled by SUSY contributions~\cite{LHSKim}, for example, by large $A$-terms. We have computed the branching ratio with the SusyBSG code~\cite{Degrassi:2007kj}, in which the next-to-next-to-leading order calculation in QCD is involved as in the calculation of Ref.~\cite{Misiak:2006zs}. \\

%%%%%%%%%  constraint from BR(Bs to mu mu)  %%%%%%%%%
\noindent
{\bf $\bullet$ Constraint on the $B_s \to \mu^+ \mu^-$ process} \\

\noindent
Diagrams mediated by neutral Higgs bosons contribute to $B_s \to \mu^+\mu^-$, and their contributions are proportional to $(\tan\beta)^6/m_A^{4}$~\cite{Bobeth:2001sq}. Since neutral Higgs boson masses are of the order of $m_Z$ in the LHS, the branching ratio of this process gives an important constraint on model parameters. Actually, the parameter region with $\tan \beta \gtrsim 10$ is almost forbidden as can be seen in the next section. We use the current experimental bound ${\rm Br} (B_s \to \mu^+ \mu^-) < 1.08 \times 10^{-8}$ at 95\%~C.L.~\cite{Bsmumu-LHC} which is given by combined analysis of results in the LHCb and the CMS experiments. \\

%%%%%%%%%  constraint from B to tau nu  %%%%%%%%%
\noindent
{\bf $\bullet$ Constraint on the $B^\pm \to \tau^\pm \nu$ process} \\

\noindent
The light charged Higgs boson $H^\pm$ contributes to the $B^\pm \to \tau^\pm \nu$ process as in the case of $B_s \to \mu^+ \mu^-$. The experimental result on the process is ${\rm Br}(B^\pm \to \tau^\pm \nu)_{\rm exp} = (1.63 \pm 0.39) \times 10^{-4}$~\cite{Barberio:2008fa}. On the other hand, the SM prediction is ${\rm Br}(B^\pm \to \tau^\pm \nu)_{\rm SM} = (1.01 \pm 0.26) \times 10^{-4}$~\cite{Akeroyd:2010qy}, somewhat smaller than the experimental value. We define their ratio as $r^{\rm exp} = {\rm Br}(B^\pm \to \tau^\pm \nu)_{\rm exp}/{\rm Br} (B^\pm \to \tau^\pm \nu)_{\rm SM}$. As a result, the deviation between experimental and theoretical results is quantitatively given by $r^{\rm exp} = 1.62 \pm 0.57$. Although the uncertainty still be large at present, the constraint is potentially important to restrict the LHS, because the additional contribution from $H^\pm$ increases the deviation~\cite{Hou:1992sy}. We have used the formula~\cite{Hall:1993gn, D'Ambrosio:2002ex, Akeroyd:2003zr, Isidori:2006pk},
\begin{eqnarray}
r^{\rm MSSM}
&\equiv&
 \frac{{\rm Br}(B^\pm \to \tau^\pm \nu)_{\rm MSSM}}
      {{\rm Br}(B^\pm \to \tau^\pm \nu)_{\rm SM}}
=
 \left\{
  1
  -
  \left( \frac{m_B^2}{m_{H^\pm}^2 }\right)^2
  \frac{\tan^2\beta}{1 + \epsilon_0 \tan\beta }
 \right\}^2,
\\
\epsilon_0
&\equiv&
 - \frac{2 \alpha_s \mu}{3\pi m_{\tilde{g}}}
 \left[
  \frac{ m_{\tilde{u}_L}^2 \ln(m_{\tilde{u}_L}^2/m_{\tilde{g}}^2) }
       {
        (m_{\tilde{g}}^2 - m_{\tilde{u}_L}^2)
        (m_{\tilde{u}_L}^2 - m_{\tilde{b}_R}^2)
       }
 +
 \frac{ m_{\tilde{b}_R}^2 \ln(m_{\tilde{b}_R}^2/m_{\tilde{g}}^2) }
      {
       (m_{\tilde{g}}^2 - m_{\tilde{b}_R}^2)
       (m_{\tilde{b}_R}^2 - m_{\tilde{u}_L}^2)
      }
\right].
\end{eqnarray}
It is worth noting that $r^{\rm MSSM}$ is always less than one. In our analysis, the experimental result with the $2\sigma$ region, $0.68 < r^{\rm exp} < 2.76$, is applied as the cut for $r^{\rm MSSM}$. \\

%%%%%%%%%  constraint from SUSY particle searches  %%%%%%%%%
\noindent
{\bf $\bullet$ Constraints from sparticle searches} \\

\noindent
In our analysis, we first derive the boundary condition at the GUT scale $M_G$ using $\tan\beta$, $\mu$, and $m_A$ at the $m_Z$ scale through the renormalization group running which is evaluated by the ISAJET~7.75 code~\cite{ISAJET}. Masses of sparticles are then obtained by running back from $M_G$ to $m_Z$ using the renormalization group running again. There are several lower bounds on the masses of sparticles as shown briefly in Table~\ref{tab:constraints}. More details on the bounds are as follows:
\begin{eqnarray*}
\text{The lightest chargino}
&:&
 m_{\chi^\pm_1} > 92.4\,\GeV \quad
 (m_{\chi^\pm_1} - m_{\chi_1} < 3\,\GeV \ \text{or} \
  m_{\tilde{\nu}} < 300\,\GeV)\\
&&
 m_{\chi^\pm_1} > 103.5\,\GeV \quad
 (\text{else})\\
\text{Charged slepton}
&:&
 m_{\tilde{\ell}_R} > 73\,\GeV \quad
 (m_{\tilde{\ell}_R} - m_{\chi_1} < 10\,\GeV)\\
&&
 m_{\tilde{e}_R} > 99.9\,\GeV, \ 
 m_{\tilde{\mu}_R} > 94.9\,\GeV, \
 m_{\tilde{\tau}_R} > 86.6\,\GeV \quad
 (\text{else})\\
\text{Sneutrino}
&:&
 m_{\tilde{\nu}} > 43\,\GeV \quad
 (m_{\tilde{\nu}} - m_{\chi_1} < 10\,\GeV)\\
&&
 m_{\tilde{\nu}} > 94\,\GeV \quad
 (\text{else})
\\
\text{Stop and sbottom}
&:&
 m_{\tilde{t}_1} > 65\,\GeV \quad
 (m_{\tilde{t}_1} - m_{\chi_1} < 10\,\GeV)\\
&&
 m_{\tilde{b}_1} > 59\,\GeV \quad
 (m_{\tilde{b}_1} - m_{\chi_1} < 10\,\GeV)\\
&&
 m_{\tilde{t}_1}, m_{\tilde{b}_1} > 95\,\GeV \quad
 (\text{else}) .
\end{eqnarray*}
Among those, the bound on $m_{\chi^\pm_1}$ is particularly important and gives a severe constraint on model parameters. This is because it determines the lower bound on $m_{1/2}$, and then the lower bound on the dark matter mass is given through the GUT relation on gaugino masses. On the other hand, the bounds on squark masses also give important constraints on model parameters because large $|A_0|$ and/or $\mu$ tend to be favored in the LHS\@. Constraints on squark mass ($m_{\tilde{q}} >379\,\GeV$) and gluino mass ($m_{\tilde{g}}>308\,\GeV$) used in the MCMC study are pre-selections. We use further cuts, $m_{\tilde{q}} >1100\,\GeV$ and $m_{\tilde{g}}>750\,\GeV$, for generated MCMC samples later in order to satisfy recent results at the ATLAS~\cite{SUSY-ATLAS} and CMS~\cite{Chatrchyan:2011zy} experiments. \\

%%%%%%%%%%  constraint from vacuum stability  %%%%%%%%%%%%%
\noindent
{\bf $\bullet$ Constraint from the vacuum stability} \\

\noindent
Too large $A$-terms gives rise to charge and color breaking~(CCB) minima. We therefore take the constraint $|A_{t}|^2 < 3 (m_{\tilde t_1}^2 + m_{\tilde t_2}^2 + m_{H_u}^2 + \mu^2)$ into account in our MCMC calculation. The constraint guarantees that the vacuum we are living is the absolute minimum even if there are CCB minima~\cite{ccb}. Since $m_{H_u}^2 + \mu^2 \sim m_Z^2$ which are much smaller than stop masses $m_{\tilde t_1}^2$ and $m_{\tilde t_2}^2$, the constraint gives an upper bound on $|A_t|$ in terms of the stop masses. The above constraint can be weaker when we also consider the case that our vacuum is metastable whose lifetime is much longer than the age of the universe. In such a case, the constraint is relaxed to be $|A_{t}|^2 + 3 \mu^2 < 7.5 (m_{\tilde t_L}^2 + m_{\tilde t_R}^2)$~\cite{Kusenko:1996jn}. We adopt, however, the first one to constrain model parameters, because it is less ambiguous than the weaker constraint. It is also worth noting that the problem of CCB minima occurs when $A_0$ is much larger than $m_0$, and such a parameter region is not favored in the view point of the naturalness. The extension of the allowed parameter region to larger $A_0$ is, hence, not so attractive. \\

%%%%%%%%%%%%%%%%%%%%%%%%%%%%%%%%%
%%%%%  subsec: LHS samples  %%%%%
%%%%%%%%%%%%%%%%%%%%%%%%%%%%%%%%%
\subsection{LHS samples}

\begin{figure}[t]
\vspace*{-20mm}
\begin{center}
\includegraphics[angle=-90,scale=0.5]{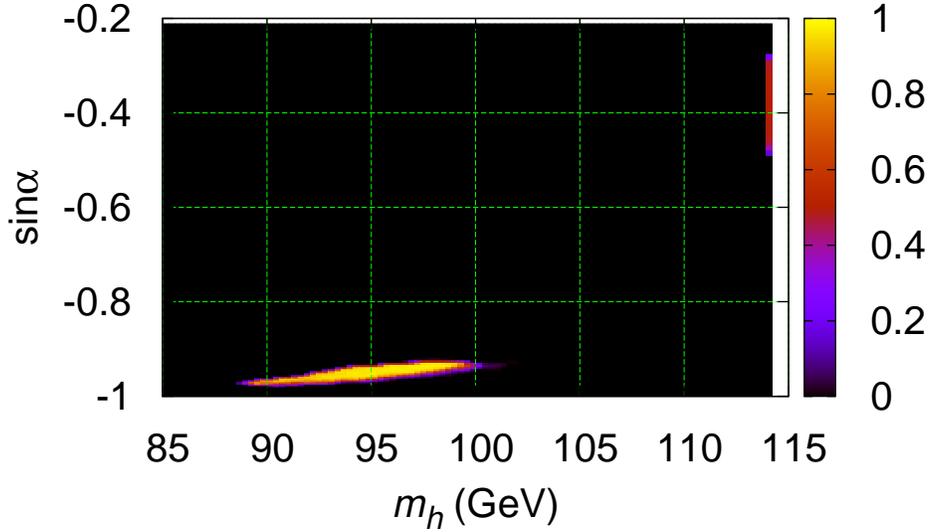}
\vspace*{-2mm}
\caption{\small Distribution of $4.0 \times 10^5$ samples in the ($m_h$, $\sin\alpha$)-plane obtained by the MCMC calculation, where $2.2 \times 10^5$ samples with $|\sin \alpha| > 0.9$ are adopted as the LHS samples in following analyses.}
\label{fig:mh_alpha}
\end{center}
\end{figure}

With the use of the constraints discussed in Sec.~\ref{sec:constraints}, we have generated $4.0 \times 10^5$ samples in the range $m_h < 114.4\,\GeV$ through the MCMC algorithm. The distribution of the samples on the ($m_h$, $\sin\alpha$)-plane is shown in Fig.~\ref{fig:mh_alpha}. Two distinct regions can be seen in the figure. In the region around $m_h = 114.4\,\GeV$, the lightest Higgs boson $h$ is composed almost of the up-type Higgs boson ${\rm Re} \, H_u^0$ because of small $|\sin \alpha|$. Thus, the physics property in the region is essentially the same as the one frequently discussed in comprehensive analyses of the CMSSM~\cite{Kane:1993td}. On the other hand, in the region with large $|\sin \alpha|$, $h$ consists dominantly of the down-type Higgs boson ${\rm Re} \, H_d^0$, which is characteristic of the LHS as discussed in Sec.~\ref{sec:LHS}. We therefore use only $2.2\times 10^5$ samples satisfying $|\sin \alpha| > 0.9$ in the next section.

\section{Physics results}
\label{sec:results}

 There are six parameters in the NUHM\@.
 However,
only two constraints
on $\Omega_{\text{DM}} h^2$ and $\text{Br}(b\to s\gamma)$
are used as Gaussian distributions
in our MCMC calculation.
 Central values of those constraints
are reproduced simultaneously not only at a point
but in a wide parameter space.
 When we see number distributions of samples
on two parameter space,
the region of large number of samples does not always means
better agreement with experimental data than other regions;
 the region of large number of samples
may mean just the large spread of the acceptable region
in the projected four parameter space.
 Therefore,
it is better to use the likelihood
in order to see the experimentally acceptable region
of the parameter space.
 The likelihood quantifies the agreement with experiments,
and we define it as the product of two Gaussian distribution functions
for $\Omega_{\text{DM}} h^2$ and $\text{Br}(b\to s\gamma)$.
 The normalization is defined so that
the set of the central values for them in Table~\ref{tab:constraints}
makes the likelihood unity.
 The larger value of the likelihood means
the better agreement with experiments.
 The maximum value of the likelihood is calculated
in each cell in two parameter space,
and we show the distributions in this section.
 In the region
where the maximum of the likelihood is close to unity
in the two parameter space,
experimental constraints can be satisfied very well
by appropriate choices of other four parameters.

%%%%%%%%%%%%%%%%%%%%%%%%%%%%%%%%%%%%%%
%%%%%  subsec: parameter region  %%%%%
%%%%%%%%%%%%%%%%%%%%%%%%%%%%%%%%%%%%%%
%%%%%%%%%%
\subsection{Parameter region}
\label{ssec:parameters}
%%%%%%%%%%

%%%%%  input parameters  %%%%%
\begin{figure}[]
\begin{center}
\includegraphics[angle=-90,width=11.2cm]{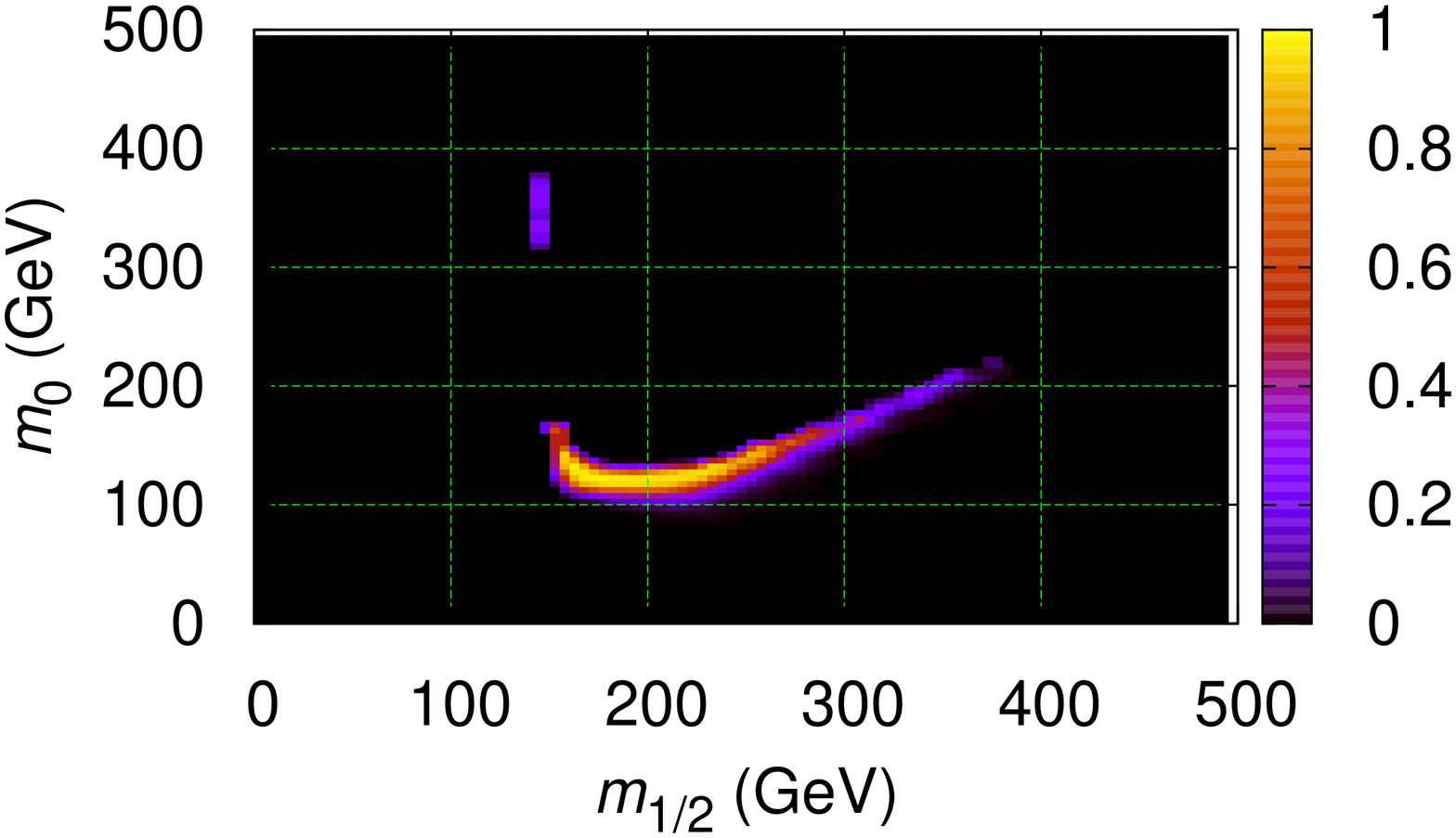}
\put(-165,-25){(a)} \\[-5mm]
\includegraphics[angle=-90,width=11.2cm]{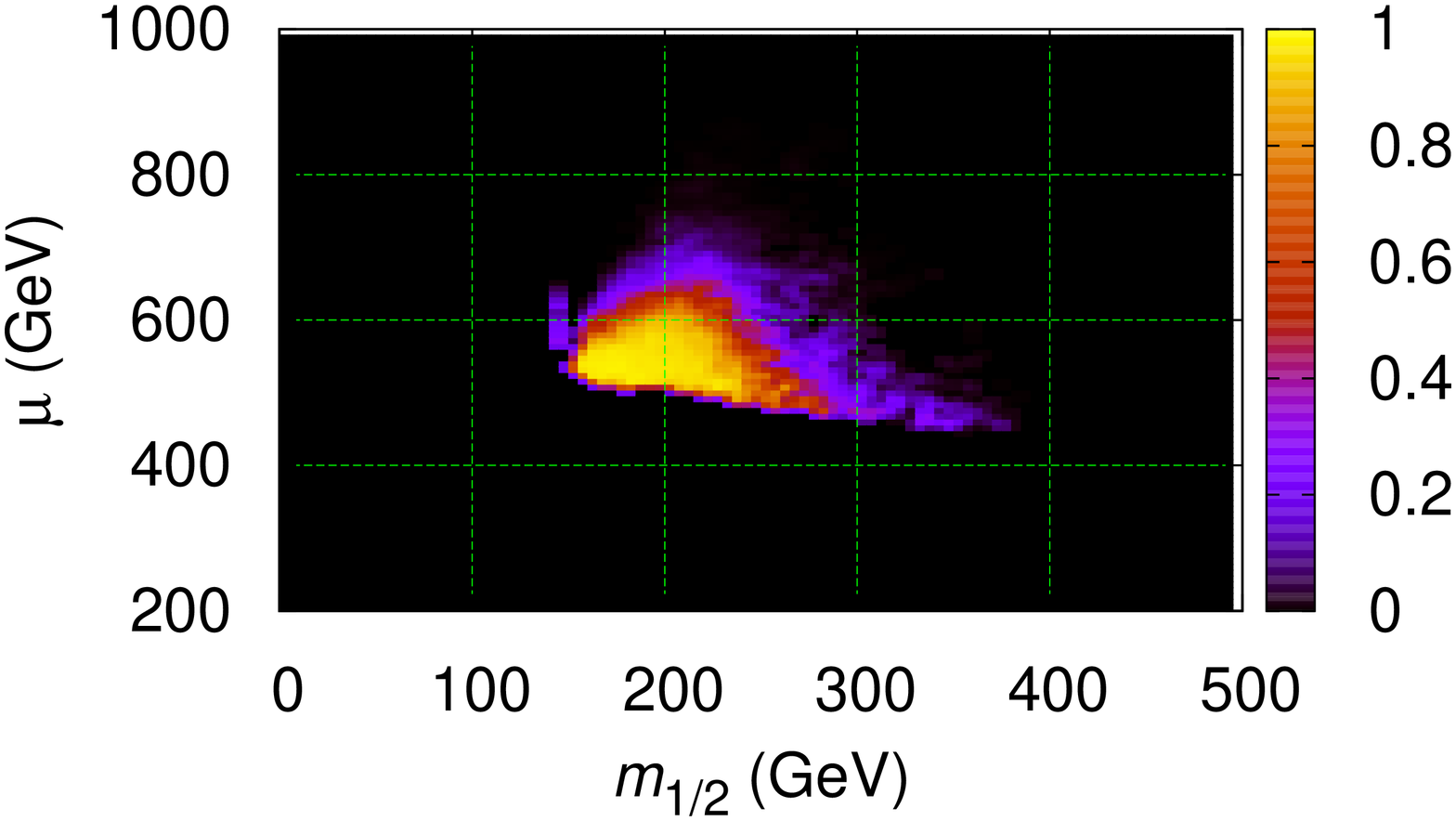}
\put(-165,-25){(b)} \\[-5mm]
\includegraphics[angle=-90,width=11.2cm]{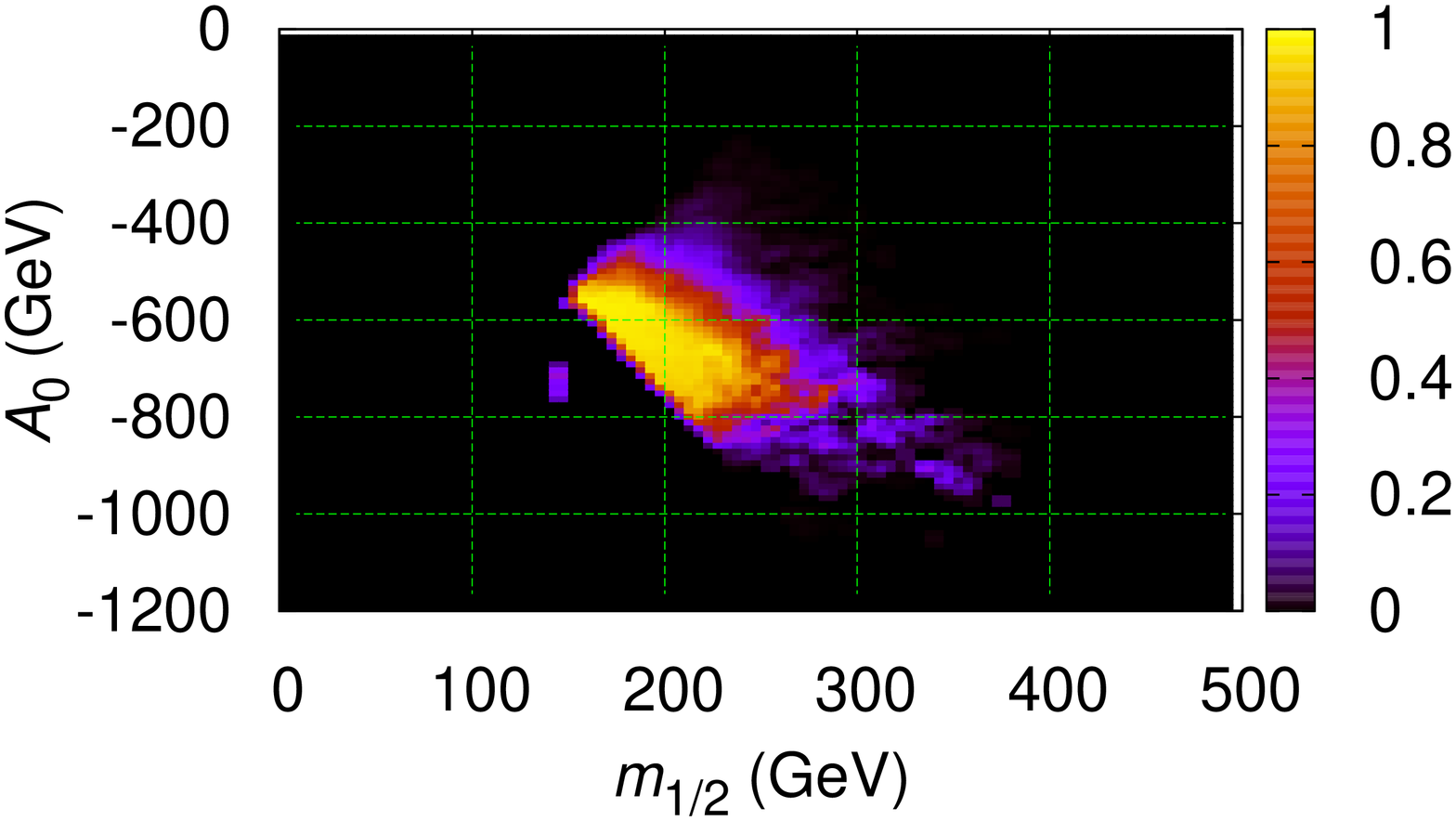}
\put(-165,-25){(c)} \\[0mm]
\caption{ \small
\label{fig:parameters}
The distribution of input parameter correlation.
The relations, (a)~$m_{1/2}\,\text{-}\,m_0$,
(b)~$m_{1/2}\,\text{-}\,\mu$,
and (c)~$m_{1/2}\,\text{-}\,A_0$
are shown, respectively.
The yellow (bright) region is of a large value of likelihood maximum in a cell.
}
\end{center}
\end{figure}

First, results for the parameter distribution in the NUHM are shown
in Fig.~\ref{fig:parameters} as the correlation of $m_{1/2}$.
The relations, (a)~$m_{1/2}\,\text{-}\,m_0$,
(b)~$m_{1/2}\,\text{-}\,\mu$, and (c)~$m_{1/2}\,\text{-}\,A_0$
are shown, respectively.
Other two parameter, $m_A$ and $\tan \beta$, are distributed in restricted regions,
$m_A \simeq 94\,\text{-}\,107\,\GeV$ and $\tan \beta \simeq 7\,\text{-}\,10$, respectively.
The yellow (bright) region is of a large value of likelihood maximum in a cell.
The contribution to $b\to s\gamma $ from charged Higgs
should be cancelled out by SUSY contribution,
and hence allowed parameters have upper bounds.
Allowed parameters in Fig.~\ref{fig:parameters}(a) are 
divided into two regions by dark matter annihilation processes.
One around $m_{1/2} = 140\,\GeV$ is pseudo-Higgs mediation region,
and the other is coannihilation region.
In the former region, the relic abundance of the LSP is governed by 
the pseudo-Higgs mediated process,
$\tilde{\chi}^0_1 \tilde{\chi}^0_1 \to A \to b \bar{b}$.
In comparison between these two regions,
the likelihood is somewhat better for the coannihilation region.

As seen in Fig.~\ref{fig:parameters}(b), 
smaller value of $\mu < 400\,\GeV$ are entirely rejected.
This is due to the direct detection constraint by XENON~100.
Since the LHS has the small Higgs masses and
a large $\tan \beta$,
direct detection rate
which is given by mediation of the lightest Higgs
(down-type one) is predicted to be large.
 For a small $\mu$,
the rate becomes too large
because of additional enhancement
due to a significant mixing
between bino and Higgsino.

In the coannihilation region, the appropriate relic abundance of the LSP 
is given by a tuning of the masses of the LSP and stau, 
i.e.\ a tuning of $m_{1/2}$ and $m_0$.
Since the relic abundance in the region is independent of the value of $\mu$,
a broad range of $\mu$ is allowed. 
In the pseudo-Higgs mediation region, 
the relic abundance depends on the value of the $\mu$ term as well as 
the mass difference between the LSP and pseudo-scalar Higgs boson mass 
because the coupling of the LSP to Higgs bosons depends on $\mu$. Meanwhile 
the relic abundance is independent of $m_0$. 
Large $\mu$ is favored in order to satisfy the constraint by direct 
detection experiments. On the other hand, as concerns little hierarchy,
too large value of $\mu$ is not favorable because light Higgs masses 
require the cancellation between very large values of $\mu$ and 
$m_{H_{d,u}}^{}$.

In Ref.~\cite{Kim:2008uh},
the authors pointed out that there is a region where the relic abundance is 
determined by the processes 
$\tilde{\chi}^0_1 \tilde{\chi}^0_1 \to hA, HA, W^\pm H^\mp, Zh$ processes 
for lower value of $\mu \simeq 300\,\GeV$.
For this small $\mu$, since the mixing between bino and Higgsino is significant,
these processes can be dominant for the control of the relic abundance of the LSP\@.
However, due to the smallness of $\mu$,
this region is excluded by the recent constraint of direct detection experiments.

In Fig.~\ref{fig:parameters}(c), the distribution of $A_0$ is shown.
The relic abundance of the LSP is independent of $A_0$,
and hence the allowed region is almost determined by
the $b \to s \gamma$ process and the condition to avoid dangerous vacua.
The large negative $A_0$ is favored by the $b \to s \gamma$ process
in order to cancel a large $H^+$ contribution
while too large $|A_0|$ leads to dangerous charge and color breaking minimum.
As a result, the allowed range of $A_0$ has upper and lower bounds.
In the coannihilation and the pseudo-Higgs mediation regions,
$|A_0|$ becomes large for a large $m_{1/2}$ and $m_0$, respectively.
This trend compensates heavy SUSY particle masses
for the contribution to $b\to s \gamma$ process.
In the pseudo-Higgs mediation region,
SUSY particles have larger masses 
than those in the coannihilation region.
Therefore, a larger $|A_0|$ is required
in the pseudo-Higgs mediation region
than that in the the coannihilation region
for $m_{1/2}\simeq 140\,\GeV$.

\begin{figure}[t]
\begin{center}
\includegraphics[angle=-90,width=10.5cm]{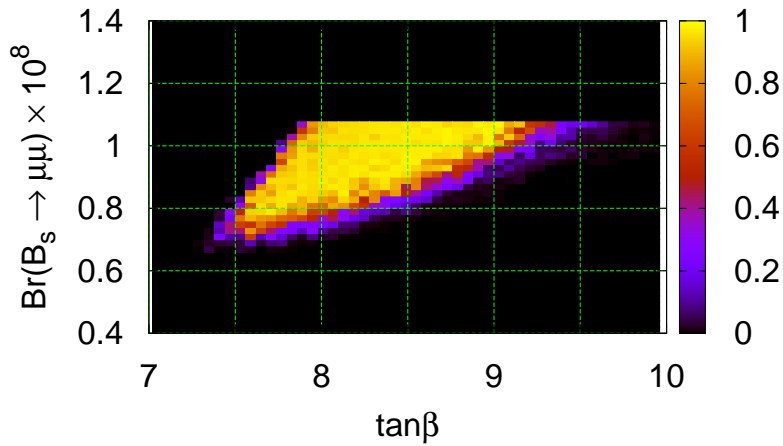}
\caption{ \small
 Distribution in a space of $\tan\beta$ and the branching ratio of 
 $B_s \to \mu^+ \mu^-$.
}
\label{fig:tanb-Bsmm}
\end{center}
\end{figure}

In addition, the $B_s \to \mu^+ \mu^-$ process gives
complementary condition to the $b \to s \gamma$ process and the relic abundance.
The branching ratio of $B_s \to \mu^+ \mu^-$ is proportional to $\tan^6 \beta / m_A^{4}$.
Hence, this process is drastically enhanced
in the LHS due to a small $m_A$.
As a result, large $\tan \beta \gtrsim 10$ is ruled out by this constraint.
On the other hand, the lower bound on $\tan \beta $ comes from the realization of the LHS~\cite{LHSBelyaev}.
The lower bound $\tan \beta \gtrsim 7$ results in
${\rm Br} (B_s \to \mu^+ \mu^-) \gtrsim 7 \times 10^{-9}$
(See Fig.~\ref{fig:tanb-Bsmm}).
Constraint on this process also excludes the region where 
lighter stop mass is very light as
$200\,\text{-}\,300\,\GeV$ due to a large $|A_0|$.
A light stop mass is favored for $b \to s \gamma$,
and however this inevitably enhances ${\rm Br} (B_s \to \mu^+ \mu^-) $.

%%%%%  input parameters  %%%%%

 Before the discussion about the LHC constraint on SUSY particles,
we mention
the distributions of Higgs boson masses.
The SM-like Higgs search at the LHC constrains the mass of the heavier 
neutral Higgs in the LHS 
because it is composed mainly of the up-type Higgs. Although stop masses 
are not large in LHS, the SM-like Higgs mass is lifted up because of
following two contributions;
one is from the large A-term which is
required to be consistent with the $b \to s \gamma$ constraint and the other
is from the off-diagonal component of the neutral Higgs mass matrix.
As a result, the mass is found to be distributed
in the range of $114\,\GeV < m_{H} < 122\,\GeV$ in our MCMC samples.
Since the LHS requires a 
large $|\sin \alpha|$ (namely, $m_A \sim m_Z$), the allowed region of 
$m_A$ is nearly the same range; $94\,\GeV \lesssim m_A \lesssim 107\,\GeV$.
Because of $m_{H^+}^2 = m_A^2 + m_W^2$ at the tree level,
$m_{H^+}$ is restricted as $122\,\GeV \lesssim m_{H^+} \lesssim 133\,\GeV$.
The ranges of $m_{H^+}$ and $\tan \beta $ are not constrained
by present results of the charged Higgs search by collider experiments.

 Finally, let us consider LHC constraints on SUSY particles. In the analysis, 
we use the NUHM as a benchmark model in which the constraint on boundary 
conditions of $m_{H_u}$ and $m_{H_d}$ are relaxed as compared with CMSSM\@. 
But the constraints on the parameter space of ($m_0$, $m_{1/2}$) would be 
exactly similar to the CMSSM case in the LHS region shown above. 
In these models, differences arise only from $m_{H_u}$ and $m_{H_d}$ values 
which can be rewritten by $m_A$ and $\mu$ at the weak scale while the masses 
of gluino and first generation of squark are nearly the same. Additionally, the 
LSP mass is also nearly the same in the LHS region shown above because the 
Higgsino component of the LSP is small. Then, the production cross section of 
SUSY particles and collider signal of these events would be no change and the 
constraint on the CMSSM parameter space~\cite{SUSY-ATLAS,Chatrchyan:2011zy} 
excludes the whole LHS region shown above.

\section{Summary and Discussions}
\label{sec:summary}

We have analyzed comprehensively the LHS of the MSSM in the framework of the NUHM,
which is the simplest one realizing the LHS in GUTs.
We have taken into account, all important constraints from collider experiments,
flavor-changing processes, and cosmological observations.
To search allowed parameter sets, we have used the MCMC method.
We have shown that there are two parameter regions consistent with these
constraints except for LHC results.
One is the coannihilation region and the other is the pseudo-Higgs mediation region.
To reconcile the $H^\pm$ contribution to the $b \to s \gamma $ process
with experimental bounds, all SUSY particles remains light, as $\lesssim 1000\,\GeV$.
Hence, we have found that 
the current LHC results have already excluded the whole LHS region in the NUHM\@. 

Finally, we discuss that some relaxation of our restriction or some extensions 
of our simplest model could have allowed LHS regions in the rest of this paper.
First, we consider the relaxation of experimental constrains
on $\Omega_{\rm DM}$ and muon $g-2$.
Other constraints are considered to be robust or
have not reduced sample points significantly.
Second, the restriction of the NUHM is relaxed.
That is, the boundary condition at the GUT scale is relaxed,
or the gauge group of the GUT is broken at an intermediate scale above the EW scale.

The relaxation of the restriction of the relic abundance of the LSP 
is expected to bring us the large number of allowed parameter sets.
It is favorable that the thermal relic abundance of the LSP is consistent with 
the WMAP observed dark matter abundance.
However, if this is not the case,
the model is not excluded unless the relic abundance is larger than the WMAP abundance.
The relic abundance can significantly be reduced for a smaller $\mu$,
since the Higgsino-bino mixing is enhanced.
The reduced abundance of the LSP
relaxes the constraint from direct detection experiments.
Then, a small $\mu$ which enhances the direct detection cross section
is allowed in this relaxation condition.
In $m_{1/2}\text{-}m_0$ space
(or $m_{\tilde{g}}\text{-}m_{\tilde{q}}$ space),
the extended allowed region of $\mu$
does not give new sample points
at the neighborhood of the coannihilation regions
where the relic abundance is independent of $\mu$.
However,
a new allowed region with $\mu \lesssim 400\,\GeV$
appears for $m_{1/2}\gtrsim 300\,\GeV$.
In this region, the LSP is significantly annihilated by 
$\tilde{\chi}^0_1 \tilde{\chi}^0_1 \to hA, HA, W^\pm H^\mp, Zh$ processes
and larger values of $m_0$ and $m_{1/2}$ are allowed.
Then, the squark and gluino masses can exceed the LHC constraints.

Next, we mention the relaxation of the muon $g-2$ constraint.
In pseudo-Higgs mediation region,
$m_{\tilde{q}}$ is extended to
$m_{\tilde{q}} \simeq 1.2\,\TeV$
by removing the $g-2$ constraint.
However,
this relaxed region is still rejected by the LHC results
because the gluino mass is light as $400\,\GeV$.
On the other hand,
the coannihilation region is extended to $m_{1/2} \simeq 450\,\GeV$.
For this upper limit, the gluino mass is about $1000\,\GeV$.
This remains as the region excluded by the LHC results.
Hence, the relaxation of the $g-2$ constraint does not bring us
new allowed parameter sets.

The relaxation of NUHM conditions changes the spectrum of the MSSM particles.
If the gauge group of the GUT is broken into some gauge group
(for example,
$\text{SU(3)}_C \times \text{SU(2)}_L
\times \text{SU(2)}_R \times \text{U(1)}_{B-L}$
of the left-right model)
at an intermediate scale above the EW scale,
the mass relation of the MSSM particles can be modified.
In the modified spectrum of the MSSM particle masses,
first and second generation squarks
are required to be heavy
because of the LHC constraints
while stop should be light
in order to have a sizable SUSY contribution
to $b \to s \gamma$.
This might be realized through the renormalization running
with a large $|A_0|$.
 Another possibility might be
that the third generation squarks
are very distinguished from first and second generations
as a boundary condition.
 Of course,
the MSSM without any boundary conditions at the GUT scale
would have enough parameter space to realize the LHS
allowed by current experimental data.
 Detailed studies of each possibilities
are beyond the scope of this paper.

Even in new parameter sets obtained by such relaxations, Higgs bosons 
should be light in the LHS and they give large contributions to
$b \to s \gamma$ and $B_s \to \mu^+ \mu^-$. Gaugino and stop should 
also be light enough to cancel a large contribution of $H^\pm$
to $b \to s \gamma$. Furthermore, the charged Higgs boson production 
is enhanced because the coupling, $g_{W^\pm H^\mp h}$, depends on 
$\cos(\beta-\alpha)$~\cite{LHSBelyaev}
(For the case with CP-violating scalar sector,
see also Ref.~\cite{Akeroyd:2003jp} and references therein).
As a result, the LHS can be 
judged in the near future by the LHC SUSY particle search, 
the LHCb search of $B_s \to \mu^+ \mu^-$ process and 
the search for $H^\pm h$ production at the LHC experiments.

{\bf Note added}:
 Updated constraint on $B_s \to \mu^+\mu^-$ at the CMS
just appeared as
$\text{Br}(B_s\to \mu^+\mu^-) < 7.7\times 10^{-9}$
at $95\%$~C.L.~\cite{Bsmumu-CMS}
during finalization of this article.
 A small region in Fig.~\ref{fig:tanb-Bsmm}
satisfies the bound,
and the LHS would be tested in near future
without assuming mass spectrum of SUSY particles.

%%%%%%%%%%%%%%%%%%%%%%%%%%%%%%%%%%%%%%%%%%%%%%%%%%%%%%
\section*{Acknowledgments}

This work is supported by the Grant-in-Aid for Science Research, Ministry of Education, Culture, Sports, Science and Technology~(MEXT), Japan [No.21740174~(S.M.), No.22244021~(S.M.\ \& M.A), and No.23740210~(H.S.)], the Grant-in-Aid for the Global COE Program Weaving Science Web beyond Particle-matter Hierarchy from the MEXT of Japan~(M.A.), World Premier International Research Center Initiative~(WPI Initiative), MEXT, Japan~(S.M.),  The Sasakawa Scientific Research Grant from the Japan Science Society~(H.S.), and the German Research Foundation (DFG) through grant BR 3954/1-1~(M.A.).
M.A.\ would like to thank Motoi Endo for useful comments on B physics.

%%%%%%%%%%%%%%%%%%%%%%%%%%%%%%%%%%%%%%%%%%%%%%%%%%%%%%


\begin{thebibliography}{99} 

%%%%%%%%%%%%%%%%%%%%%%%%
%%%%% Introduction %%%%%
%%%%%%%%%%%%%%%%%%%%%%%%

%%%%  WMAP 5-year mean: Omega_c h^2 = 0.1099 +- 0.0062  %%%%%%%%
%%%%  table 1 %%%%%%
%%%%  used in our analysis  %%%%
\bibitem{WMAP}
%\bibitem{Komatsu:2008hk}
E.~Komatsu {\it et al.}  [WMAP Collaboration],
%``Five-Year Wilkinson Microwave Anisotropy Probe (WMAP) Observations: Cosmological Interpretation,''
Astrophys.\ J.\ Suppl.\  {\bf 180}, 330 (2009).
% [arXiv:0803.0547 [astro-ph]].

%%%%  WMAP 7-year mean: Omega_c h^2 = 0.1120 +- 0.0056  %%%%%%%%
%%%%  table 1 %%%%%%
%%%%  not used in our analysis  %%%%
\bibitem{recentWMAP}
%\bibitem{Komatsu:2010fb}
E.~Komatsu {\it et al.}  [WMAP Collaboration],
%``Seven-Year Wilkinson Microwave Anisotropy Probe (WMAP) Observations: Cosmological Interpretation,''
Astrophys.\ J.\ Suppl.\  {\bf 192}, 18 (2011).
% [arXiv:1001.4538 [astro-ph.CO]].

\bibitem{DMreview}
For reviews, see for instance, 
G.~Jungman, M.~Kamionkowski and K.~Griest,
Phys.\ Rept.\  {\bf 267}, 195 (1996);
L.~Bergstrom,
Rept.\ Prog.\ Phys.\  {\bf 63}, 793 (2000);
C.~Munoz,
Int.\ J.\ Mod.\ Phys.\ A {\bf 19}, 3093 (2004);
G.~Bertone, D.~Hooper and J.~Silk,
Phys.\ Rept.\  {\bf 405}, 279 (2005).

\bibitem{BookDrees}
See, for example,
M.~Drees, R.~M.~Godbole and P.~Roy,
{\it Theory and phenomenology of sparticles},
(World Scientific, 2004).

\bibitem{LEPSM}
%\bibitem{Barate:2003sz}
R.~Barate {\it et al.} [LEP Working Group for Higgs boson searches and ALEPH, DELPHI, L3, and OPAL Collaborations],
%``Search for the standard model Higgs boson at LEP,''
Phys.\ Lett.\  B {\bf 565}, 61 (2003).
%[arXiv:hep-ex/0306033].

\bibitem{radiative}
%\bibitem{Okada:1990vk}
Y.~Okada, M.~Yamaguchi and T.~Yanagida,
%``Upper bound of the lightest Higgs boson mass in the minimal supersymmetric standard model,''
Prog.\ Theor.\ Phys.\  {\bf 85} (1991) 1;
%\bibitem{Haber:1990aw}
H.~E.~Haber and R.~Hempfling,
%``Can the mass of the lightest Higgs boson of the minimal supersymmetric model be larger than m(Z)?,''
Phys.\ Rev.\ Lett.\  {\bf 66} (1991) 1815;
%\bibitem{Ellis:1990nz}
J.~R.~Ellis, G.~Ridolfi and F.~Zwirner,
%``Radiative corrections to the masses of supersymmetric Higgs bosons,''
Phys.\ Lett.\  B {\bf 257} (1991) 83.


%\bibitem{LHS1}
\bibitem{LHSKane}
G.~L.~Kane, T.~T.~Wang, B.~D.~Nelson and L.~T.~Wang,
Phys.\ Rev.\  D {\bf 71}, 035006 (2005).

\bibitem{LHSBelyaev}
A.~Belyaev, Q.~H.~Cao, D.~Nomura, K.~Tobe and C.~P.~Yuan,
Phys.\ Rev.\ Lett.\  {\bf 100}, 061801 (2008).

\bibitem{LHSDrees}
M.~Drees,
Phys.\ Rev.\  D {\bf 71}, 115006 (2005).
%[arXiv:hep-ph/0502075].

\bibitem{LHSKim}
S.~G.~Kim, N.~Maekawa, A.~Matsuzaki, K.~Sakurai, A.~I.~Sanda and T.~Yoshikawa,
Phys.\ Rev.\  D {\bf 74}, 115016 (2006).

\bibitem{Asano:2007gv}
M.~Asano, S.~Matsumoto, M.~Senami and H.~Sugiyama,
%``Neutralino Dark Matter in Light Higgs Boson Scenario,''
Phys.\ Lett.\  B {\bf 663}, 330 (2008).
%[arXiv:0711.3950 [hep-ph]].

\bibitem{LHSdummy}
  %\bibitem{LHSBottino1}
  A.~Bottino, F.~Donato, N.~Fornengo and S.~Scopel,
  Phys.\ Rev.\  D {\bf 63}, 125003 (2001);
  %\bibitem{LHSBottino2}
  A.~Bottino, N.~Fornengo and S.~Scopel,
  Nucl.\ Phys.\  B {\bf 608}, 461 (2001).
  
\bibitem{LEPSUSY}
S.~Schael {\it et al.}  [LEP Working Group for Higgs boson searches],
Eur.\ Phys.\ J.\  C {\bf 47}, 547 (2006).

\bibitem{Asano:2009kj}
M.~Asano, S.~Matsumoto, M.~Senami and H.~Sugiyama,
%``CDMS II result and Light Higgs Boson Scenario of the MSSM,''
JHEP {\bf 1007}, 013 (2010).
%[arXiv:0912.5361 [hep-ph]].

\bibitem{Asano:2008gi}
M.~Asano, T.~Kubo, S.~Matsumoto and M.~Senami,
Phys.\ Rev.\  D {\bf 80}, 095017 (2009).
%[arXiv:0807.4922 [hep-ph]].

\bibitem{Kim:2008uh}
  S.~G.~Kim, N.~Maekawa, K.~I.~Nagao, K.~Sakurai and T.~Yoshikawa,
  Phys.\ Rev.\  D {\bf 78} (2008) 075010.

%\cite{Funakubo:2009eg}
\bibitem{Funakubo:2009eg}
  K.~Funakubo and E.~Senaha,
  %``Electroweak phase transition, critical bubbles and sphaleron
%decoupling condition in the MSSM,''
  Phys.\ Rev.\ D {\bf 79} (2009) 115024.
%  [arXiv:0905.2022 [hep-ph]].
  %%CITATION = ARXIV:0905.2022;%%

%\cite{Eriksson:2008cx}
\bibitem{Eriksson:2008cx}
  D.~Eriksson, F.~Mahmoudi and O.~Stal,
  %``Charged Higgs bosons in Minimal Supersymmetry: Updated constraints
%and experimental prospects,''
  JHEP {\bf 0811} (2008) 035;
%  [arXiv:0808.3551 [hep-ph]].
  %%CITATION = ARXIV:0808.3551;%%
%\cite{Das:2010kb}
%\bibitem{Das:2010kb}
  D.~Das, A.~Goudelis and Y.~Mambrini,
  %``Exploring SUSY light Higgs boson scenarios via dark matter
%experiments,''
  JCAP {\bf 1012} (2010) 018;
%  [arXiv:1007.4812 [hep-ph]].
  %%CITATION = ARXIV:1007.4812;%%
%\cite{Heinemeyer:2011aa}
%\bibitem{Heinemeyer:2011aa}
  S.~Heinemeyer, O.~Stal and G.~Weiglein,
  %``Interpreting the LHC Higgs Search Results in the MSSM,''
  arXiv:1112.3026 [hep-ph];
  %%CITATION = ARXIV:1112.3026;%%
%\cite{Bottino:2011xv}
%\bibitem{Bottino:2011xv}
  A.~Bottino, N.~Fornengo and S.~Scopel,
  %``Phenomenology of light neutralinos in view of recent results at the
%CERN Large Hadron Collider,''
  arXiv:1112.5666 [hep-ph].
  %%CITATION = ARXIV:1112.5666;%%


\bibitem{Gambino:2001ew}
P.~Gambino and M.~Misiak,
%``Quark mass effects in anti-B ---> X(s gamma),''
Nucl.\ Phys.\  B {\bf 611}, 338 (2001).
% [arXiv:hep-ph/0104034].

\bibitem{Aprile:2011hi}
E.~Aprile {\it et al.}  [XENON100 Collaboration],
%``Dark Matter Results from 100 Live Days of XENON100 Data,''
Phys.\ Rev.\ Lett.\  {\bf 107}, 131302 (2011).
% arXiv:1104.2549 [astro-ph.CO].


\bibitem{SUSY-ATLAS}
  G.~Aad {\it et al.}  [Atlas Collaboration],
  %``Search for new phenomena in final states with large jet multiplicities and missing transverse momentum using sqrt(s)=7 TeV pp collisions with the ATLAS detector,''
  JHEP {\bf 1111} (2011) 099;
%  [arXiv:1110.2299 [hep-ex]]
  %%CITATION = ARXIV:1110.2299;%%
%%%  SUSY search at ATLAS, L = 1.04 fb^(-1)  %%%%
%%%  https://twiki.cern.ch/twiki/bin/view/AtlasPublic/AtlasResultsEPS2011  %%%
%I.~Vivarelli,
%talk at Europhysics Conference on High-Energy Physics 2011.
%
P.~de~Jong,
talk at Hadron Collider Physics Symposium 2011.


%\cite{Chatrchyan:2011zy}
\bibitem{Chatrchyan:2011zy}
%%  SUSY search at CMS, L = 1.14 fb^(-1)  %%%%
%%  https://twiki.cern.ch/twiki/bin/view/CMSPublic/PhysicsResults  %%%
  S.~Chatrchyan {\it et al.}  [CMS Collaboration],
  %``Search for Supersymmetry at the LHC in Events with Jets and Missing
  %Transverse Energy,''
  Phys.\ Rev.\ Lett.\ {\bf 107}, 221804 (2011);
%  arXiv:1109.2352 [hep-ex].
  %%CITATION = ARXIV:1109.2352;%%
%
%
S.A.~Koay,
talk at Hadron Collider Physics Symposium 2011.


%%%%%%%%%%%%%%%
%%%%% LHS %%%%%
%%%%%%%%%%%%%%%

%%%%%%% mSUGRA/CMSSM %%%%%%%%
\bibitem{CMSSM}
%\cite{Chamseddine:1982jx}
%\bibitem{Chamseddine:1982jx}
  A.~H.~Chamseddine, R.~L.~Arnowitt and P.~Nath,
  %``Locally Supersymmetric Grand Unification,''
  Phys.\ Rev.\ Lett.\  {\bf 49}, 970 (1982);
  %%CITATION = PRLTA,49,970;%%
%
%
%\cite{Nath:1983aw}
%\bibitem{Nath:1983aw}
  P.~Nath, R.~L.~Arnowitt and A.~H.~Chamseddine,
  %``Gauge Hierarchy In Supergravity Guts,''
  Nucl.\ Phys.\  B {\bf 227}, 121 (1983);
  %%CITATION = NUPHA,B227,121;%%
%
%
%\cite{Hall:1983iz}
%\bibitem{Hall:1983iz}
  L.~J.~Hall, J.~D.~Lykken and S.~Weinberg,
  %``Supergravity As The Messenger Of Supersymmetry Breaking,''
  Phys.\ Rev.\  D {\bf 27}, 2359 (1983);
  %%CITATION = PHRVA,D27,2359;%%
%
%
%\cite{Arnowitt:1992aq}
%\bibitem{Arnowitt:1992aq}
  R.~L.~Arnowitt and P.~Nath,
  %``SUSY mass spectrum in SU(5) supergravity grand unification,''
  Phys.\ Rev.\ Lett.\  {\bf 69}, 725 (1992);
  %%CITATION = PRLTA,69,725;%%
%
%
%\cite{Ross:1992tz}
%\bibitem{Ross:1992tz}
  G.~G.~Ross and R.~G.~Roberts,
  %``Minimal supersymmetric unification predictions,''
  Nucl.\ Phys.\  B {\bf 377}, 571 (1992);
  %%CITATION = NUPHA,B377,571;%%
%
%
%\cite{Barger:1993gh}
%\bibitem{Barger:1993gh}
  V.~D.~Barger, M.~S.~Berger and P.~Ohmann,
  %``The Supersymmetric particle spectrum,''
  Phys.\ Rev.\  D {\bf 49}, 4908 (1994).
%  [arXiv:hep-ph/9311269].
  %%CITATION = PHRVA,D49,4908;%%


\bibitem{Kane:1993td}
G.~L.~Kane, C.~F.~Kolda, L.~Roszkowski and J.~D.~Wells,
%``Study of constrained minimal supersymmetry,''
Phys.\ Rev.\  D {\bf 49}, 6173 (1994).
% [arXiv:hep-ph/9312272]

%%%%%%%%%%%%%%%%%%%%%
%%%%% Framework %%%%%
%%%%%%%%%%%%%%%%%%%%%

\bibitem{recipes}
See e.g.,
W.H.~Press {\it et al.},
{\it Numerical Recipes 3rd Edition},
(Cambridge University Press).

\bibitem{Nakamura:2010zzi}
K.~Nakamura  [Particle Data Group],
%``Review of particle physics,''
J.\ Phys.\ G {\bf 37} (2010) 075021.

%%%%%  BR( b to s gamma ) < (352 +- 23 +- 9)*10^(-6)  %%%%%%%%
%%%%%  table 120  %%%%%%
%%%%%  used in our analysis  %%%%%
\bibitem{Barberio:2008fa}
E.~Barberio {\it et al.}  [Heavy Flavor Averaging Group],
%``Averages of $b-$hadron and $c-$hadron Properties at the End of 2007,''
arXiv:0808.1297 [hep-ex].

%%%  upper bound on BR( Bs to mu mu ) %%%%
%%%   CMS L=1.14 fb^(-1): 1.9*10^(-8) (95% CL)  %%%%
%%%  LHCb L=0.34 fb^(-1): 1.5*10^(-8) (95% CL)  %%%%
%%%  CMS+LHCb: 1.08*10^(-8) (95% CL)  %%%%
\bibitem{Bsmumu-LHC}
%``Search for the rare decay $B^{0}_{s}\to \mu^{+}\mu^{-}$ at the LHC with the CMS and LHCb experimentsCombination of LHC results of the search for $B_s\to\mu^+\mu^-$ decays,''
CMS-PAS-BPH-11-019;
LHCb-CONF-2011-047;
%
%\cite{Chatrchyan:2011kr}
%\bibitem{Chatrchyan:2011kr}
  S.~Chatrchyan {\it et al.}  [CMS Collaboration],
  %``Search for B(s) and B to dimuon decays in pp collisions at 7 TeV,''
  Phys.\ Rev.\ Lett.\ {\bf 107}, 191802 (2011);
%  arXiv:1107.5834 [hep-ex].
  %%CITATION = ARXIV:1107.5834;%%
%
%\cite{LHCb:2011ac}
%\bibitem{LHCb:2011ac}
 R.~Aaij {\it et al.}  [LHCb Collaboration],
 %``Search for the rare decays Bs -> mu+ mu- and B0 -> mu+ mu-,''
 Phys.\ Lett.\  B {\bf 708}, 55 (2012).
% [arXiv:1112.1600 [hep-ex]].
 %%CITATION = PHLTA,B708,55;%%


\bibitem{Akeroyd:2010qy}
A.~G.~Akeroyd and F.~Mahmoudi,
%``Measuring V_ub and probing SUSY with double ratios of purely leptonic decays of B and D mesons,''
JHEP {\bf 1010}, 038 (2010).
% [arXiv:1007.2757 [hep-ph]]

\bibitem{LEPSUSYweb}
LEP2 SUSY Working Group Webpage,
\verb$http://lepsusy.web.cern.ch/lepsusy$

\bibitem{Abdallah:2003xe}
J.~Abdallah {\it et al.}  [DELPHI Collaboration],
Eur.\ Phys.\ J.\  C {\bf 31}, 421 (2003).

\bibitem{:2007ww}
V.~M.~Abazov {\it et al.}  [D0 Collaboration],
Phys.\ Lett.\  B {\bf 660}, 449 (2008).

%%%%%%  muon g-2  %%%%%%%%  
%\cite{Hagiwara:2011af}
\bibitem{Hagiwara:2011af}
K.~Hagiwara, R.~Liao, A.~D.~Martin, D.~Nomura and T.~Teubner,
%``(g-2)_mu and alpha(M_Z^2) re-evaluated using new precise data,''
J.\ Phys.\ G {\bf 38}, 085003 (2011).
% [arXiv:1105.3149 [hep-ph]].

%%%%  updated constraint BR(b to s gamma) < (355 +- 24 +- 9)*10^(-6)  %%%%
%%%%  table 125  %%%%%
%%%%  not used in our analysis  %%%%
\bibitem{Asner:2010qj}
D.~Asner {\it et al.}  [Heavy Flavor Averaging Group],
%``Averages of b-hadron, c-hadron, and $\tau-lepton Properties,''
arXiv:1010.1589 [hep-ex].

%%%%%%  micrOMEGAs  %%%%%%%%
\bibitem{Belanger:2008sj}
G.~Belanger, F.~Boudjema, A.~Pukhov and A.~Semenov,
Comput.\ Phys.\ Commun.\  {\bf 180} (2009) 747.

\bibitem{Kamionkowski:2008vw}
M.~Kamionkowski and S.~M.~Koushiappas,
%``Galactic Substructure and Direct Detection of Dark Matter,''
Phys.\ Rev.\  D {\bf 77}, 103509 (2008).
% [arXiv:0801.3269 [astro-ph]].

\bibitem{Ohki:2008ff}
  H.~Ohki {\it et al.},
  Phys.\ Rev.\  D {\bf 78} (2008) 054502;
%
%\bibitem{Ohki:2009mt}
%  H.~Ohki {\it et al.},
  %``Nucleon sigma term and strange quark content in 2+1-flavor QCD with
  %dynamical overlap fermions,''
  PoS {\bf LAT2009}, 124 (2009).
%  [arXiv:0910.3271 [hep-lat]].
  %%CITATION = POSCI,LAT2009,124;%%
  
%%%%%  b to s gamma in the SM  %%%%
\bibitem{Becher:2006pu}
T.~Becher and M.~Neubert,
Phys.\ Rev.\ Lett.\  {\bf 98}, 022003 (2007).

%%%%%  SusyBSG  %%%%%
\bibitem{Degrassi:2007kj}
G.~Degrassi, P.~Gambino and P.~Slavich,
Comput.\ Phys.\ Commun.\  {\bf 179} (2008) 759.

\bibitem{Misiak:2006zs}
M.~Misiak {\it et al.},
%``The first estimate of B(anti-B --> X/s gamma) at O(alpha(s)**2),''
Phys.\ Rev.\ Lett.\  {\bf 98}, 022002 (2007).
%[arXiv:hep-ph/0609232]

\bibitem{Bobeth:2001sq}
C.~Bobeth, T.~Ewerth, F.~Kruger and J.~Urban,
%``Analysis of neutral Higgs boson contributions to the decays $\bar{B}$($s^{)} \to \ell^{+} \ell^{-}$ and $\bar{B} \to K \ell^{+} \ell^{-}$,''
Phys.\ Rev.\  D {\bf 64}, 074014 (2001).
% [arXiv:hep-ph/0104284]

\bibitem{Hou:1992sy}
W.~S.~Hou,
%``Enhanced charged Higgs boson effects in B- $\to$ tau anti-neutrino, mu anti-neutrino and b $\to$ tau anti-neutrino + X,''
Phys.\ Rev.\  D {\bf 48}, 2342 (1993).

\bibitem{Hall:1993gn}
L.~J.~Hall, R.~Rattazzi and U.~Sarid,
%``The Top quark mass in supersymmetric SO(10) unification,''
Phys.\ Rev.\  D {\bf 50}, 7048 (1994).
% [arXiv:hep-ph/9306309]

\bibitem{D'Ambrosio:2002ex}
G.~D'Ambrosio, G.~F.~Giudice, G.~Isidori and A.~Strumia,
%``Minimal flavour violation: An effective field theory approach,''
Nucl.\ Phys.\  B {\bf 645}, 155 (2002).
% [arXiv:hep-ph/0207036]

%\cite{Akeroyd:2003zr}
\bibitem{Akeroyd:2003zr}
  A.~G.~Akeroyd and S.~Recksiegel,
  %``The effect of H+- on B+- --> tau+- nu/tau and B+- --> mu+- nu/mu,''
  J.\ Phys.\ G {\bf 29}, 2311 (2003).
%  [arXiv:hep-ph/0306037].
  %%CITATION = JPHGB,G29,2311;%%

\bibitem{Isidori:2006pk}
G.~Isidori and P.~Paradisi,
%``Hints of large tan(beta) in flavour physics,''
Phys.\ Lett.\  B {\bf 639}, 499 (2006).
% [arXiv:hep-ph/0605012]

\bibitem{ISAJET}
F.~E.~Paige, S.~D.~Protopopescu, H.~Baer and X.~Tata,
arXiv:hep-ph/0312045.

\bibitem{ccb}
J.~M.~Frere, D.~R.~T.~Jones and S.~Raby,
%``Fermion Masses And Induction Of The Weak Scale By Supergravity,''
Nucl.\ Phys.\  B {\bf 222}, 11 (1983);
%\bibitem{Claudson:1983et}
M.~Claudson, L.~J.~Hall and I.~Hinchliffe,
%``Low-Energy Supergravity: False Vacua And Vacuous Predictions,''
Nucl.\ Phys.\  B {\bf 228}, 501 (1983).

\bibitem{Kusenko:1996jn}
A.~Kusenko, P.~Langacker and G.~Segre,
%``Phase Transitions and Vacuum Tunneling Into Charge and Color Breaking Minima in the MSSM,''
Phys.\ Rev.\  D {\bf 54}, 5824 (1996).
% [arXiv:hep-ph/9602414]

%\cite{Akeroyd:2003jp}
\bibitem{Akeroyd:2003jp}
  A.~G.~Akeroyd,
  %``Searching for a very light Higgs boson at the Tevatron,''
  Phys.\ Rev.\  D {\bf 68}, 077701 (2003).
%  [arXiv:hep-ph/0306045].
  %%CITATION = PHRVA,D68,077701;%%


\bibitem{Bsmumu-CMS}
U.~Langenegger [the CMS collabolation],
talk at CERN, 28 Feb.\ 2012,\\
\verb#http://indico.cern.ch/conferenceDisplay.py?confId=178806#

\end{thebibliography}
\end{document}